\documentclass[
preprint,
showpacs,preprintnumbers,
bibnotes,
 amsmath,amssymb,
 aps,
 pra,
superscriptaddress,
longbibliography,
]{revtex4-1}

\usepackage[version=3]{mhchem} 
\usepackage[dvipsnames]{xcolor}
\usepackage{textcomp}
\usepackage{graphicx}
\usepackage{amsmath}
\usepackage{fixmath}
\usepackage{amsfonts}
\usepackage{amssymb}

\newcommand{\micron}{\hbox{\textmu}\text{m}}

\graphicspath{ {./figures/} }

\begin{document}

\title{
Spin--valley dynamics in alloy-based transition metal dichalcogenide heterobilayers 
}

\author{Vasily Kravtsov}
\email{Corresponding author: vasily.kravtsov@metalab.ifmo.ru}
\affiliation{Department of Physics and Engineering, ITMO University, Saint Petersburg 197101, Russia}
\author{Aleksey D. Liubomirov}
\author{Roman V. Cherbunin}
\affiliation
{Saint Petersburg State University, ul. Ulyanovskaya 1, Saint Petersburg 198504, Russia}
\author{Alessandro Catanzaro}
\author{Armando Genco}
\author{Daniel Gillard}
\author{Evgeny M. Alexeev}
\affiliation{Department of Physics and Astronomy, University of Sheffield, Sheffield S3 7RH, UK}
\author{Tatiana Ivanova}
\author{Ekaterina Khestanova}
\affiliation{Department of Physics and Engineering, ITMO University, Saint Petersburg 197101, Russia}
\author{Ivan A. Shelykh}
\affiliation{Department of Physics and Engineering, ITMO University, Saint Petersburg 197101, Russia}
\affiliation{Science Institute, University of Iceland, Dunhagi 3, IS-107, Reykjavik, Iceland}
\author{Ivan V. Iorsh}
\affiliation{Department of Physics and Engineering, ITMO University, Saint Petersburg 197101, Russia}
\author{Alexander I. Tartakovskii}
\affiliation
{Department of Physics and Astronomy, University of Sheffield, Sheffield S3 7RH, UK}
\author{Maurice S. Skolnick}
\author{Dmitry N. Krizhanovskii}
\email{Corresponding author: d.krizhanovskii@sheffield.ac.uk}
\affiliation{Department of Physics and Engineering, ITMO University, Saint Petersburg 197101, Russia}
\affiliation
{Department of Physics and Astronomy, University of Sheffield, Sheffield S3 7RH, UK}

\date{\today}



\begin{abstract}
\noindent 
\textbf{
Van der Waals heterobilayers based on 2D transition metal dichalcogenides have been recently shown to support robust and long-lived valley polarization for potential valleytronic applications.
However, the role of the band structure and alignment of the constituent layers in the underlying dynamics remains largely unexplored.
Here we study spin--valley relaxation dynamics in heterobilayers with different band structures engineered via the use of alloyed monolayer semiconductors.
Through a combination of time-resolved Kerr rotation spectroscopic measurements and theoretical modelling for Mo$_{1-x}$W$_{x}$Se$_2$/WSe$_2$ samples with different chemical compositions and stacking angles, we uncover the roles of interlayer exciton recombination and charge carrier spin depolarization in the overall valley dynamics.
Our results provide insights into the microscopic spin--valley polarization mechanisms in van der Waals heterostructures for the development of future 2D valleytronic devices.
}
\end{abstract}

\maketitle

\noindent Two-dimensional transition metal dichalcogenides (TMDs) possess two sets of optically addressable helicity-selective valleys (K$^+$/K$^-$) and thus provide a very promising material platform for the development of valleytronic devices~\cite{Yao2008,Mak2012,Schaibley2016,Vitale2018}.
For practical applications, valley polarization has to persist on extended time scales.
However, valley relaxation dynamics in monolayer (ML) TMDs are often severely limited by the efficient exchange interaction and fast radiative recombination of direct intralayer excitons~\cite{Zhu2014,DalConte2015,Plechinger2017,Wang2018}.

On the other hand, heterobilayers (HBLs) consisting of two vertically stacked different TMD monolayers can exhibit significantly longer valley lifetimes owing to the ultrafast separation of electrons and holes between the layers via tunneling and suppression of the fast valley relaxation channels~\cite{Rivera2016,Rivera2018}.
Recent studies have demonstrated valley relaxation times of resident holes in such structures up to the microsecond scale~\cite{Kim2017,Jin2018Science,Jiang2018}, opening possibilities for the electrical control of valley properties~\cite{Ciarrocchi2019}.
However, the underlying microscopic mechanisms of spin and valley relaxation, including interactions between different excitonic species and charge carriers, are not yet fully understood.
Moreover, open questions remain regarding the effects of the TMD chemical composition, stacking angle, and band alignment~\cite{Jin2018}.

Atomically thin layers of TMD alloys, such as Mo$_{1-x}$W$_{x}$Y$_2$ (Y = Se, S), provide an attractive means to explore the connection between valley relaxation processes and intrinsic material properties defined by the band structure~\cite{Dumcenco2013,Ye2017,Liang2019}.
Through a continuous tuning of the W/Mo relative concentration $x$, monolayer alloys with controllable band gap, exciton transition frequency, and spin--orbit splitting in the conduction and valence bands can be obtained~\cite{Ke2015,Wang2015}.
Therefore, heterostructures based on monolayer TMD alloys~\cite{Zhang2019,Zi2019} in principle provide effective control on the band alignment and allow access to the continuous transition towards TMD homobilayers~\cite{Liu2014}, where pronounced hybridization effects can lead to rich exciton physics~\cite{Gerber2019,Brem2020} and associated valley dynamics~\cite{Scuri2020}.

In this work, we investigate the spin--valley relaxation dynamics in TMD alloy-based heterobilayers with controllable chemical composition and stacking angle.
By probing the spin--valley polarization on the direct exciton resonances of both constituent layers, we uncover the coupled dynamics of spin- and valley-polarized interlayer excitons and resident electrons, and extract the corresponding interaction strength.
We introduce a model that takes into account different tunneling rates of electrons and holes in the heterobilayer and fully describes the observed dynamics and their temperature dependencies.
Our results provide insight into the fundamental mechanisms of valley relaxation in van der Waals heterostructures and demonstrate control on valley polarization dynamics via alloying.

\begin{figure}[t]
	\includegraphics[width=0.65\columnwidth]{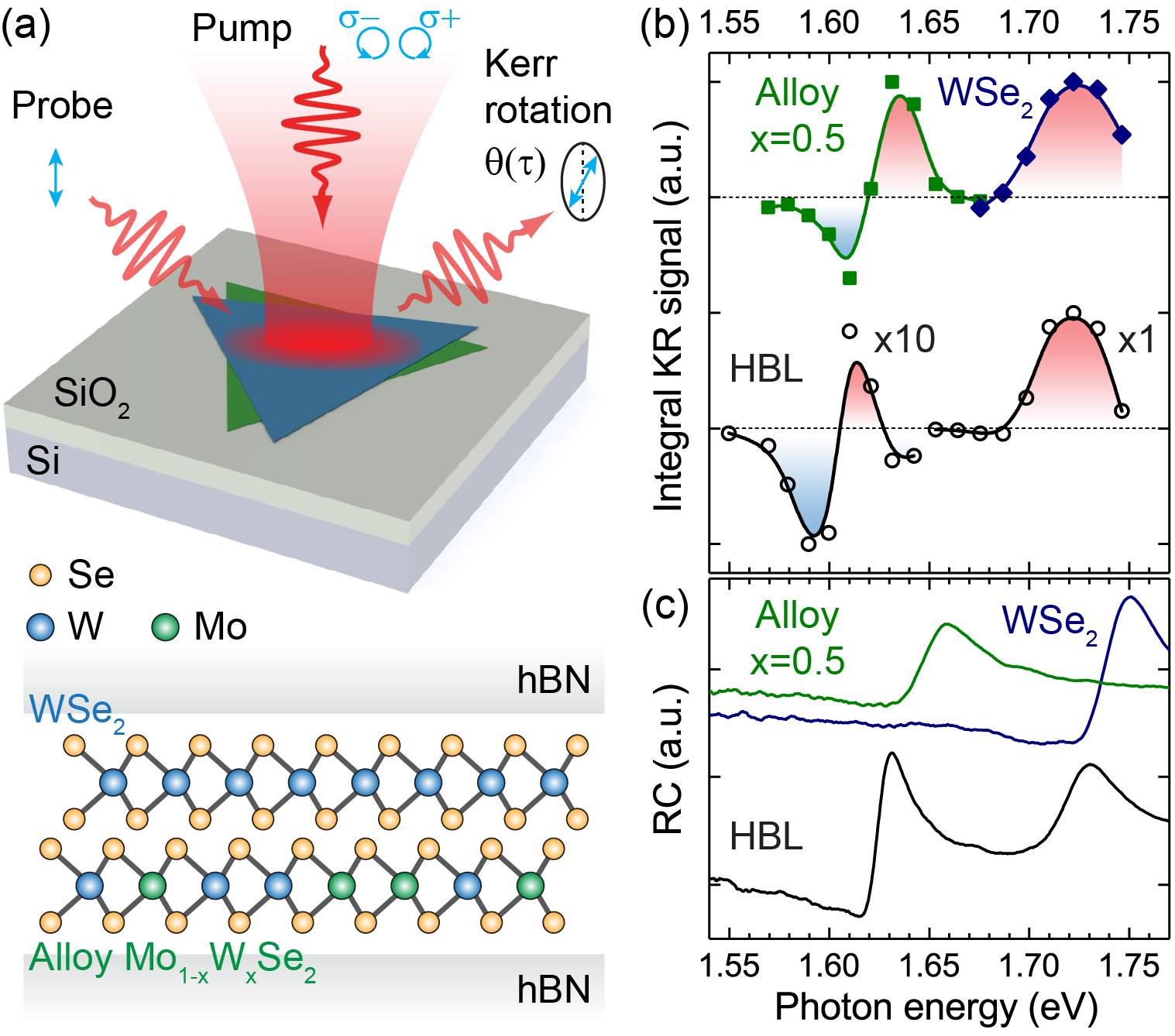}
	\caption{\textbf{Kerr rotation spectroscopy of alloy-based TMD heterobilayers.} (a) Top: schematic of the experiment, with circularly polarized pump and linearly polarized probe pulses. Bottom: Schematic of the sample, consisting of a Mo$_{1-x}$W$_{x}$Se$_2$/WSe$_2$ heterobilayer encapsulated in few-layer hBN. (b) Time-integrated and normalized Kerr rotation signal measured as a function of the excitation/detection frequency for Mo$_{0.5}$W$_{0.5}$Se$_2$ monolayer (green squares), WSe$_2$ monolayer (blue diamonds), and Mo$_{0.5}$W$_{0.5}$Se$_2$/WSe$_2$ heterobilayer (open black circles). The lines are guides to the eye. (c) Reflectivity contrast spectra for the  Mo$_{0.5}$W$_{0.5}$Se$_2$ monolayer (green), WSe$_2$ monolayer (blue), and Mo$_{0.5}$W$_{0.5}$Se$_2$/WSe$_2$ heterobilayer (black).}
	\label{fig:Setup}
\end{figure}

We experimentally study spin--valley dynamics in hBN/Mo$_{1-x}$W$_{x}$Se$_2$/WSe$_2$/hBN stacks (Fig.~\ref{fig:Setup}a, bottom panel) with time-resolved Kerr rotation (KR) micro-spectroscopy.
The sample fabrication procedures and measurement technique are described in Methods.
As illustrated in the top panel of Fig.~\ref{fig:Setup}a, degenerate circularly polarized pump and linearly polarized probe pulses with a controlled delay are focused onto the sample, and the rotation angle of the probe polarization plane is measured in the reflection geometry as a function of pump-probe delay.

First, we consider the Kerr rotation response in a Mo$_{0.5}$W$_{0.5}$Se$_2$/WSe$_2$ heterobilayer.
The KR spectrum is plotted in Fig.~\ref{fig:Setup}b as the normalized time-integrated KR angle for varying pump/probe pulse center wavelength.
For the HBL region (black open circles), two Fano-shaped resonances are observed: one at $\sim 1.72$~eV corresponding to the A exciton resonance in monolayer WSe$_2$ and the other at $\sim 1.61$~eV corresponding to that in Mo$_{0.5}$W$_{0.5}$Se$_2$, with the KR responses of the WSe$_2$ and alloy monolayers shown by green and blue symbols, respectively.
The corresponding reflectivity contrast spectra are presented in (c).
The HBL resonances are slightly red-shifted with respect to the ML resonances due to the modified dielectric environment~\cite{Raja2017,Kunstmann2018}.
Additionally, the KR resonances are red-shifted with respect to the reflectivity contrast spectral features, which suggests a contribution from trions.

\begin{figure*}[t]
	\includegraphics[width=\columnwidth]{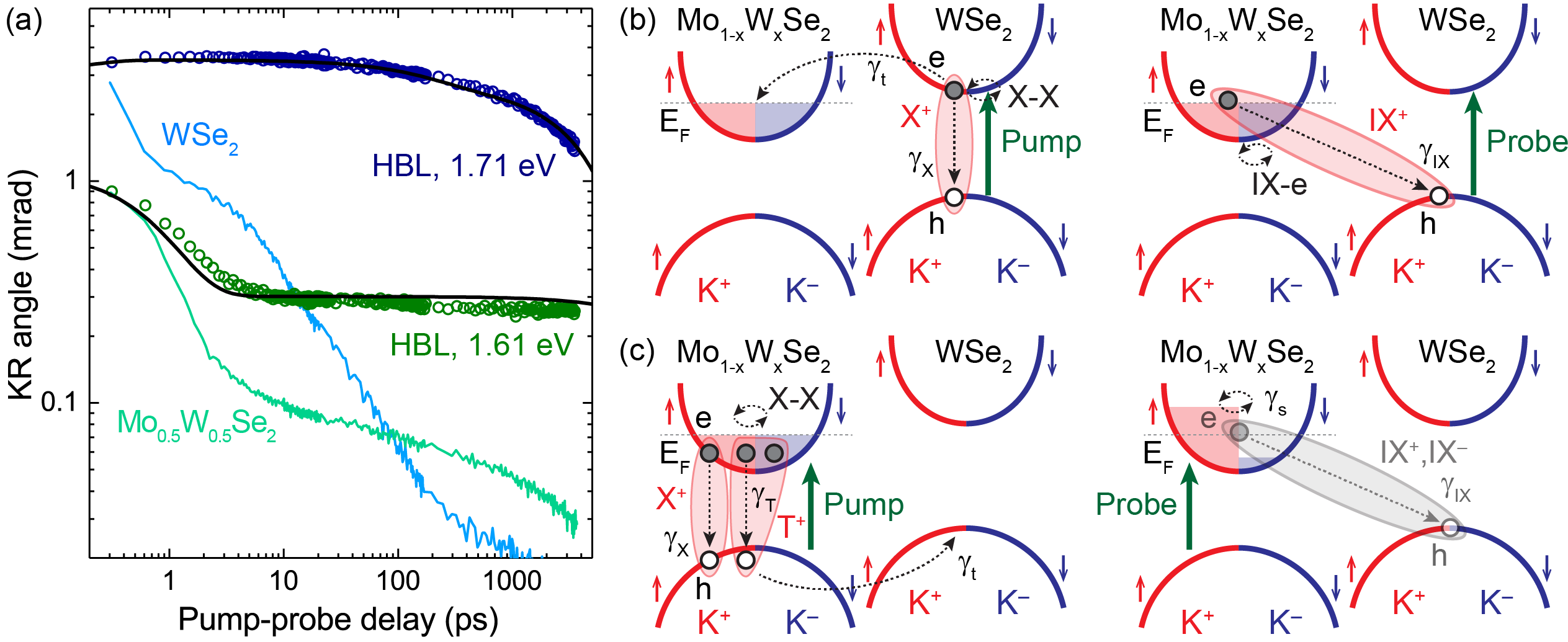}
	\caption{\textbf{Spin--valley relaxation dynamics in $x=0.5$ alloy-based heterobilayers.} (a) Kerr rotation kinetics measured on a Mo$_{0.5}$W$_{0.5}$Se$_2$/WSe$_2$ heterobilayer at $\hbar\omega = 1.71$~eV (W resonance, dark blue circles) and $\hbar\omega = 1.61$~eV (M resonance, dark green circles), together with corresponding curves for monolayer WSe$_2$ (light blue) and Mo$_{0.5}$W$_{0.5}$Se$_2$ (light green), plotted in a log--log scale. Black solid curves are model calculations. (b) Schematic diagram of the valley dynamics in a heterobilayer for excitation/detection at the W resonance: direct excitons in WSe$_2$ are quickly separated between the layers by electron tunneling (left panel), forming interlayer excitons that depolarize by interaction with resident electrons in the alloy and recombine (right panel). (c) Schematic diagram for the M resonance: competition between hole tunneling and exciton/trion depolarization and recombination (left) creates spin-polarized resident electrons in the alloy, while interlayer exciton population and valley polarization are small (right). $\gamma_\mathrm{X}, \gamma_\mathrm{T}, \gamma_\mathrm{IX}$ are recombination rates of intralayer excitons, trions, and interlayer excitons, respectively; $\gamma_\mathrm{t}, \gamma_\mathrm{s}$ are tunneling and electron spin depolarization rates, respectively.}
	\label{fig:Dynamics}
\end{figure*}

For the following discussion, we denote the observed resonances as M (exciton in the alloy ML) and W (exciton in the WSe$_2$ ML).
The dynamics of the KR response for the M and W resonances in the HBL are shown in Fig.~\ref{fig:Dynamics}a in a log--log scale, and exhibit qualitatively different behaviors.
For the W resonance (dark blue circles), the KR angle increases for $\sim 1$~ps, followed by decay on an intermediate $100-1000$~ps scale and further slower few-ns dynamics, while for the M resonance (dark green circles) it shows a fast ($\sim 1$~ps) decrease followed by a slow decay on a 100 ns scale.
The KR kinetics taken on the WSe$_2$ and alloy ML regions in the same sample are shown as light blue and light green curves, respectively.

To describe the observed KR dynamics, we introduce a simple kinetic model (see Supplementary Information for details).
The experimentally measured Kerr rotation angle $\theta_\mathrm{KR}$, which reflects the instantaneous spin--valley polarization in the sample, is defined by the difference in pump-induced reflectivity changes $\delta r_+$ and $\delta r_-$ for excitons in K$^{+}$ and K$^{-}$ valleys, respectively.
In our case it is mainly due to the valley-dependent exciton frequency shift $\delta\omega_{\pm}$: $\theta_\mathrm{KR} = -\mathrm{Im}\left [(\delta r_+ - \delta r_-)/2r\right] \propto (\delta\omega_+ - \delta\omega_-).$
The frequency shift arises from the interaction of intralayer excitons (DX) with pump-induced populations of intralayer and interlayer (IX) excitons and free charge carriers (C).

Our samples are slightly \textit{n}-doped, which is supported by the existence of long-lived KR signals and sizeable trion photoluminescence intensity in the alloy ML, as well as gate-dependent measurements on similar structures.
The Mo$_{1-x}$W$_{x}$Se$_2$/WSe$_2$ HBL is of type II band alignment~\cite{Zhang2019}, where the alloy conduction band (CB) is lower in energy than that of WSe$_2$, as shown in Fig.~\ref{fig:Dynamics}b.
This results in a finite density of resident electrons in the CB of the Mo$_{0.5}$W$_{0.5}$Se$_2$ layer and absence of resident holes in the WSe$_2$ layer, defining different valley dynamics for the W and M resonances in the HBL.

For the W resonance, the KR angle is expressed as
$$
\theta_\mathrm{KR} \propto V_{DD}(n_{DX}^{+} - n_{DX}^{-}) + V_{DI}(n_{IX}^{+} - n_{IX}^{-}),
$$
where coefficients $V_{DD}$, $V_{DI}$ represent the strength of the intralayer--intralayer and intralayer--interlayer exciton interactions, respectively, and $n_{DX,IX}^{+,-}$ are the populations of intralayer and interlayer excitons with spin $+1$ and $-1$.
After the pump-induced DX formation in the K$^{+}$ valley of WSe$_2$, electrons quickly ($< 100$~fs) tunnel into the alloy layer~\cite{Hong2014,Zhu2017}, efficiently transferring the valley polarization to interlayer excitons, as schematically illustrated in Fig.~\ref{fig:Dynamics}b.
Since $V_{DI}$ can be higher than $V_{DD}$, the KR angle can exhibit a short initial increase, as observed in the experimental data in (a).

After the electrons have tunneled into the alloy, the slower KR dynamics is determined by the IX depolarization and recombination.
While the exchange interaction is suppressed for IX, valley depolarization can still happen via interaction with un-polarized resident electrons in the CB of the alloy layer.
Considering the recombination rate $\gamma_{IX}$, density of resident electrons $n_e$, and IX--electron interaction strength $V_{IC}$ \cite{Harmon2010}, we write kinetic equations for the population dynamics:
\begin{align*}
    &\dot{m}_e(t)=-m_e(t)\tilde{n}_{IX}(0)e^{-\gamma_{IX}t}V_{IC}+m_{IX}(t)n_eV_{IC}, \\
    &\dot{m}_{IX}(t)=-m_{IX}(t)n_eV_{IC}+m_e(t)\tilde{n}_{IX}(0)e^{-\gamma_{IX}t}V_{IC},
\end{align*}
where $m_{IX}(t), m_e(t)$ indicate the valley polarization for interlayer excitons and resident electrons, respectively, and $\tilde{n}_{IX}(0)$ is the IX population created initially through electron tunneling. 
The solution (see Supplementary Information) shows two different time scales for the KR dynamics with rates $\gamma_{IX}$ for IX recombination and $\gamma_{XC}=n_eV_{IC}$ for carrier-induced depolarization.
By fitting the KR data for the W resonance with this model (Fig.~\ref{fig:Dynamics}a, black curve), we extract the electron concentration $n_e \sim 10^{11}$~cm$^{-2}$ and IX--electron interaction strength $V_{IC} \sim 10^{-2}~\mu\mathrm{eV}~\mu \mathrm{m}^2$.
The valley polarization dynamics here is limited by IX recombination, which is different from the case of \textit{p}-doped samples, where resident holes can maintain polarization on the microsecond scale~\cite{Kim2017,Jin2018Science}.

\begin{figure*}[t]
	\includegraphics[width=\columnwidth]{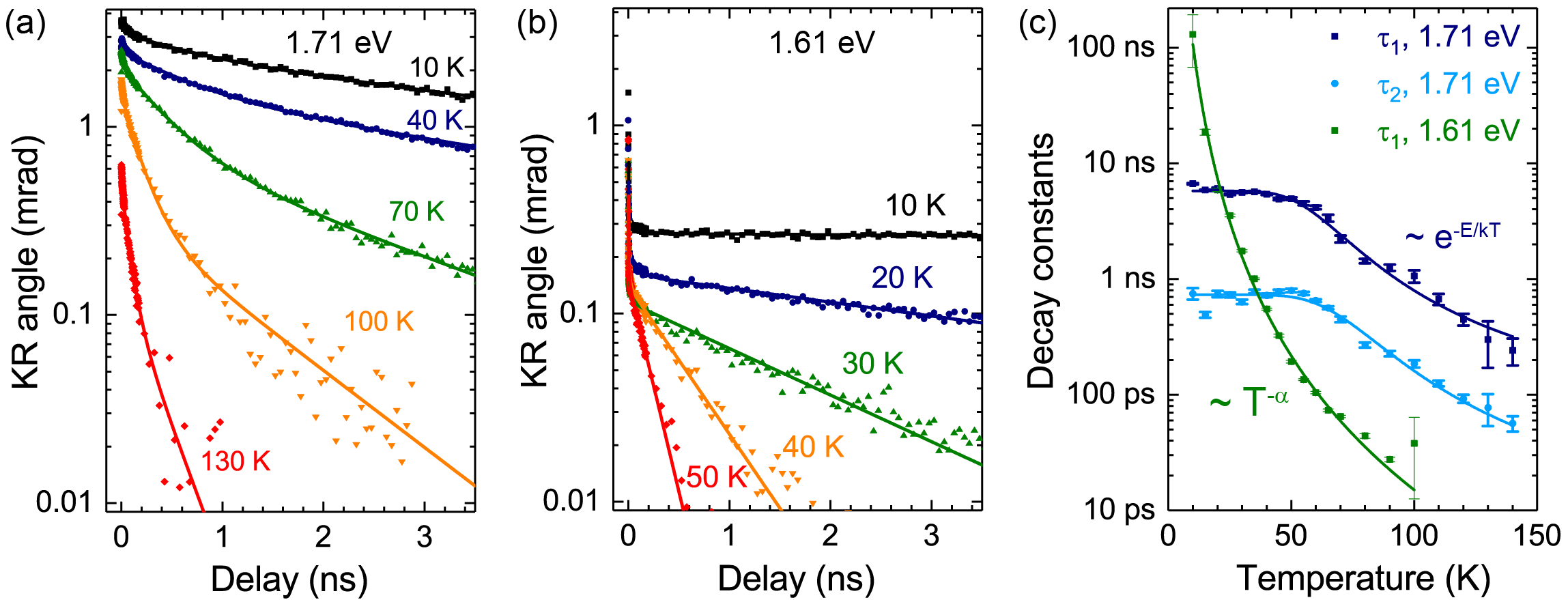}
	\caption{\textbf{Temperature dependence of the spin--valley dynamics in Mo$_{0.5}$W$_{0.5}$Se$_2$/WSe$_2$ heterobilayers.} (a) Kerr rotation kinetics measured on a heterobilayer sample for $\hbar\omega = 1.71$~eV at selected sample temperatures, indicated in the plot. (b) Kerr rotation kinetics for $\hbar\omega = 1.61$~eV measured at selected temperatures. (c) Time constants, extracted from fits to the experimental data, as functions of sample temperature (symbols), together with model curves (lines). The largest time constant for the $\hbar\omega = 1.61$~eV resonance shows an inverse power dependence on temperature (green), while the two time constants for the $\hbar\omega = 1.71$~eV resonance exhibit a thermally-activated behavior (dark and light blue).}
	\label{fig:Temperature}
\end{figure*}

When exciting the M resonance in the HBL (Fig.~\ref{fig:Dynamics}c), direct excitons formed in the alloy layer experience very fast valley depolarization~\cite{Ye2017}.
The hole tunneling rate is estimated to be an order of magnitude lower than that for the electrons in WSe$_2$ (see Supplementary Information), resulting in a fast initial valley depolarization on $\sim 1$~ps scale as observed in Fig.~\ref{fig:Dynamics}a.
The following long $\sim 100$~ns KR signal decay is due to the net valley polarization of resident electrons in the CB of the alloy layer.
A possible mechanism for this polarization involves resonant excitation of trions, consisting of an exciton in the K$^{+}$ valley bound to a resident electron in the K$^{-}$ valley, followed by their fast (100s of fs) depolarization and subsequent recombination~\cite{Hao2017}, which effectively creates a spin-polarized cloud of resident electrons~\cite{Anghel2018} in the K$^{+}$ valley, as schematically illustrated in Fig.~\ref{fig:Dynamics}c.

Due to the competition between hole tunneling and recombination of excitons/trions, the population of interlayer excitons is small in this case, and their valley polarization is low, making the IX--electron interaction inefficient.
Then the slow KR signal dynamics is dominated by the spin depolarization of the resident electrons in the CB of the alloy layer.
We note that in comparison to the HBL case, the KR signal in the monolayer Mo$_{0.5}$W$_{0.5}$Se$_2$ exhibits a more pronounced initial decay (Fig.~\ref{fig:Dynamics}a, light green curve), associated with more efficient exciton/trion recombination in the absence of hole tunneling, and dynamics on a $\sim 1$~ns scale, which can be related to localized states protected from interaction-induced depolarization~\cite{Ersfeld2019}.


Our model for the KR dynamics near the M and W resonances in the HBL is further supported by corresponding temperature dependencies.
KR kinetics measured at selected sample temperatures are shown in Fig.~\ref{fig:Temperature}a,b (symbols) for the W and M resonances, respectively, together with corresponding fit curves (lines).
The time constants for the slow valley dynamics, extracted from the fits, are plotted in Fig.~\ref{fig:Temperature}c as functions of temperature and reveal qualitatively different behavior for the two resonances.
For the M resonance (green symbols), the long decay time decreases very rapidly with temperature and exhibits an inverse power dependence $\tau \propto T^{-\alpha}$ with $\alpha \sim 3.7$ (green curve), which suggests its association with the spin depolarization processes of the resident electrons in the CB of Mo$_{0.5}$W$_{0.5}$Se$_2$ via Elliott--Yafet type mechanisms.

In contrast, for the W resonance, the extracted time constants shown in Fig.~\ref{fig:Temperature}c with dark and light blue symbols exhibit a thermally-activated behavior, with little variation for temperatures $T < 60$~K.
This behavior can be described~\cite{Volmer2017} via temperature-dependent relaxation rates $\gamma (T) = \gamma_0 + \gamma_{\infty}\operatorname{exp}(-E_\mathrm{a}/k_\mathrm{B}T)$, where $\gamma_0, \gamma_{\infty}$ are the zero- and high-temperature decay rates, $E_\mathrm{a}$ is the activation energy, and $k_\mathrm{B}$ is the Boltzmann constant.
The corresponding fits to the experimental data are shown in Fig.~\ref{fig:Temperature}c with dark and light blue curves.
For the largest time constant $\tau_1$ (dark blue), corresponding to the interlayer exciton recombination process, we extract an activation energy of $E_{a1} \sim 30$~meV, while for the intermediate time constant $\tau_2$ (light blue), corresponding to the IX--electron scattering process, we extract a similar albeit slightly higher activation energy of $E_{a2} \sim 40$~meV.
These values are close to the range of optical phonon frequencies in TMD monolayers~\cite{Peng2016}, suggesting that both the IX recombination and IX--e interaction are phonon-activated processes.

\begin{figure}[t]
	\includegraphics[width=0.65\columnwidth]{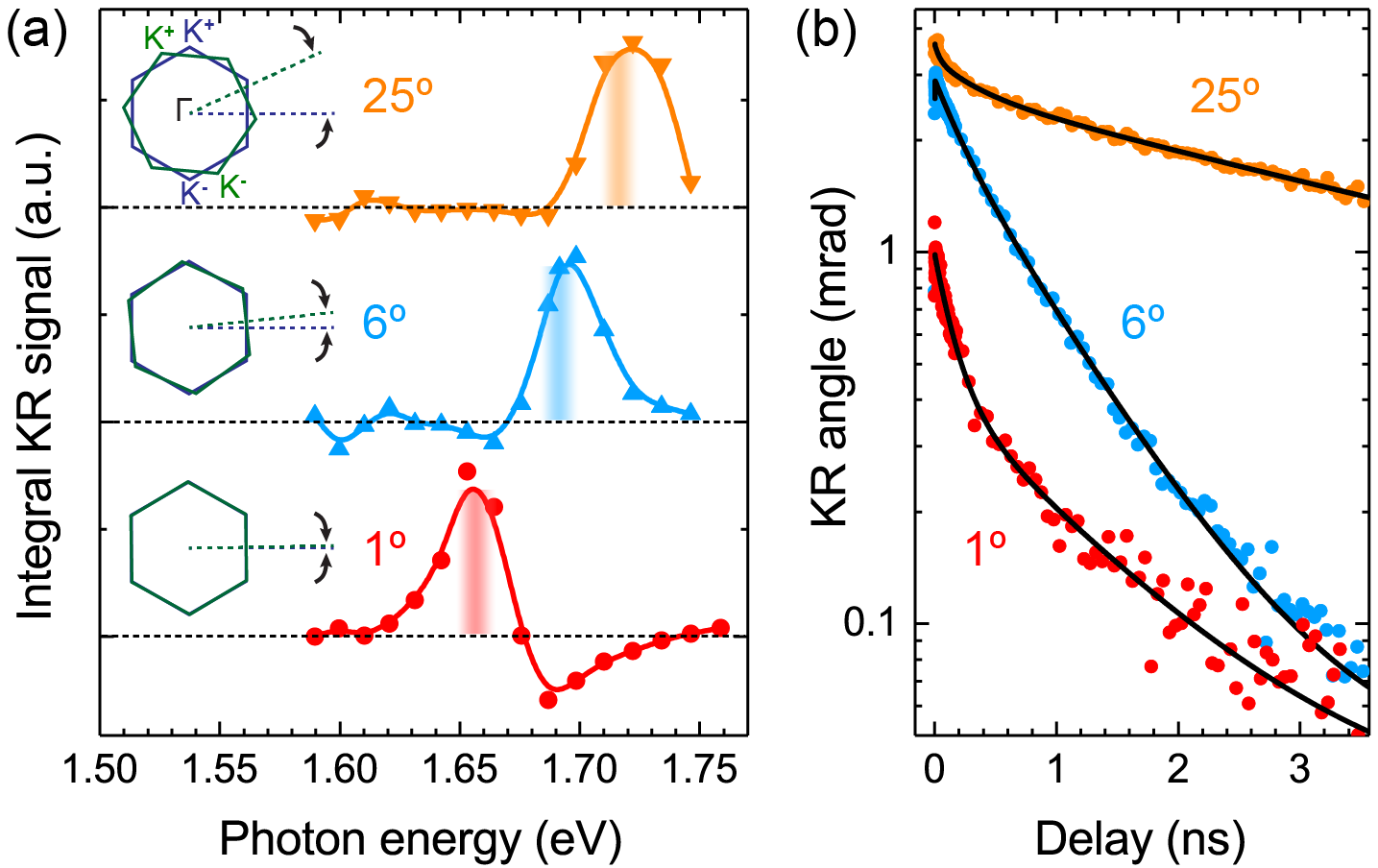}
	\caption{\textbf{Spin--valley dynamics in Mo$_{0.5}$W$_{0.5}$Se$_2$/WSe$_2$ heterobilayers with different stacking angles.} (a) Time-integrated and normalized Kerr rotation signal as a function of excitation/detection frequency for heterobilayers with stacking angle = 25$^\circ$ (orange symbols), 6$^\circ$ (blue), and 1$^\circ$ (red). The lines are guides to the eye. Inset cartoons illustrate the corresponding mismatch in the crystal momentum space. (b) Kerr rotation kinetics for heterobilayers with 3 selected stacking angles (symbols), together with fits (curves).}
	\label{fig:Angle}
\end{figure}

Using the described model of spin--valley relaxation, we proceed to explore the spin--valley dynamics in alloy-based heterobilayer samples with different band structures.
We first compare the results of KR measurements on Mo$_{0.5}$W$_{0.5}$Se$_2$/WSe$_2$ heterostructures with different stacking angles.
Fig.~\ref{fig:Angle} shows time-integrated KR spectra (a) and corresponding time traces (b) measured on the W resonance at a temperature of 10~K for three different samples with stacking angles of $\theta \sim 25^{\circ}$ (orange color), $6^{\circ}$ (blue), and $1^{\circ}$ (red).
The stacking angle, which defines the relative orientation between the crystal axes of the two constituent layers (see insets in Fig.~\ref{fig:Angle}a) and accordingly the momentum mismatch between their K$^+$/K$^-$ points in the Brillouin zone, is controlled during the fabrication process and precisely measured via polarization-resolved second harmonic generation spectroscopy as described in Methods and Supplementary Information.

With decreasing stacking angle, in Fig.~\ref{fig:Angle}a we observe a continuous spectral shift of the Fano-like W resonance peak in the KR spectra to lower photon energies, accompanied by a slight spectral broadening.
For the almost fully aligned structure ($1^{\circ}$), the resonance is red shifted by $\sim 70$~meV with respect to the misaligned one ($25^{\circ}$), with similar shifts observed in the reflectivity contrast spectra (see Supplementary Information).
This behavior likely arises from the perturbations of the excitonic bands due to hybridization effects, as has been recently predicted~\cite{Ruiz2019} and observed experimentally in homo- and hetero-bilayers of 2D semiconductors~\cite{Gerber2019,Alexeev2019}.
For small stacking angles, the momentum mismatch between the K$^+$/K$^-$ valleys in the two layers is minimized, and interlayer charge tunneling starts playing a role, which causes angle-dependent direct exciton red shift and broadening~\cite{Alexeev2019}.
The spectral shifts and broadening observed in our Mo$_{0.5}$W$_{0.5}$Se$_2$/WSe$_2$ samples suggest that, while there is still a significant difference between the conduction band edges of Mo$_{0.5}$W$_{0.5}$Se$_2$ and WSe$_2$ monolayers~\cite{Zhang2019}, hybridization with the interlayer exciton~\cite{Ruiz2019} is still possible via the delocalized wavefunction of the valence band holes~\cite{Nayak2017}.

The experimentally measured KR kinetics are shown in Fig.~\ref{fig:Angle}b.
According to the model, we extract two time constants for the IX recombination rate $\gamma_{IX}$ and IX depolarization rate $\gamma_{XC}$ due to interaction with resident electrons in the CB of the alloy monolayer.
The corresponding IX recombination time $\tau_{IX} = \gamma_{IX}^{-1}$ decreases significantly at small stacking angles from 4.8~ns for $\theta = 25^\circ$ to 0.8~ns and 1.2~ns for $\theta = 6^\circ$ and $1^\circ$, respectively.
This is due to faster radiative IX recombination for smaller momentum mismatch between the K$^+$/K$^-$ valleys in the two layers and better overlap between the electron and hole wavefunctions, as well as hybridization with short-lived direct excitons in WSe$_2$.
A similar general trend, although not as pronounced, is observed for the valley depolarization time $\tau_{XC} = \gamma_{XC}^{-1}$, which is reduced from 340~ps for $\theta = 25^\circ$ to 240~ps and 160~ps for $\theta = 6^\circ$ and $1^\circ$, respectively.
Assuming similar resident electron density in the three studied samples, which is reasonable because they are assembled from the same monolayers and protected from environment by hBN, these results imply that the IX--electron interaction strength is increased for smaller stacking angles.

\begin{figure*}[t]
	\includegraphics[width=\columnwidth]{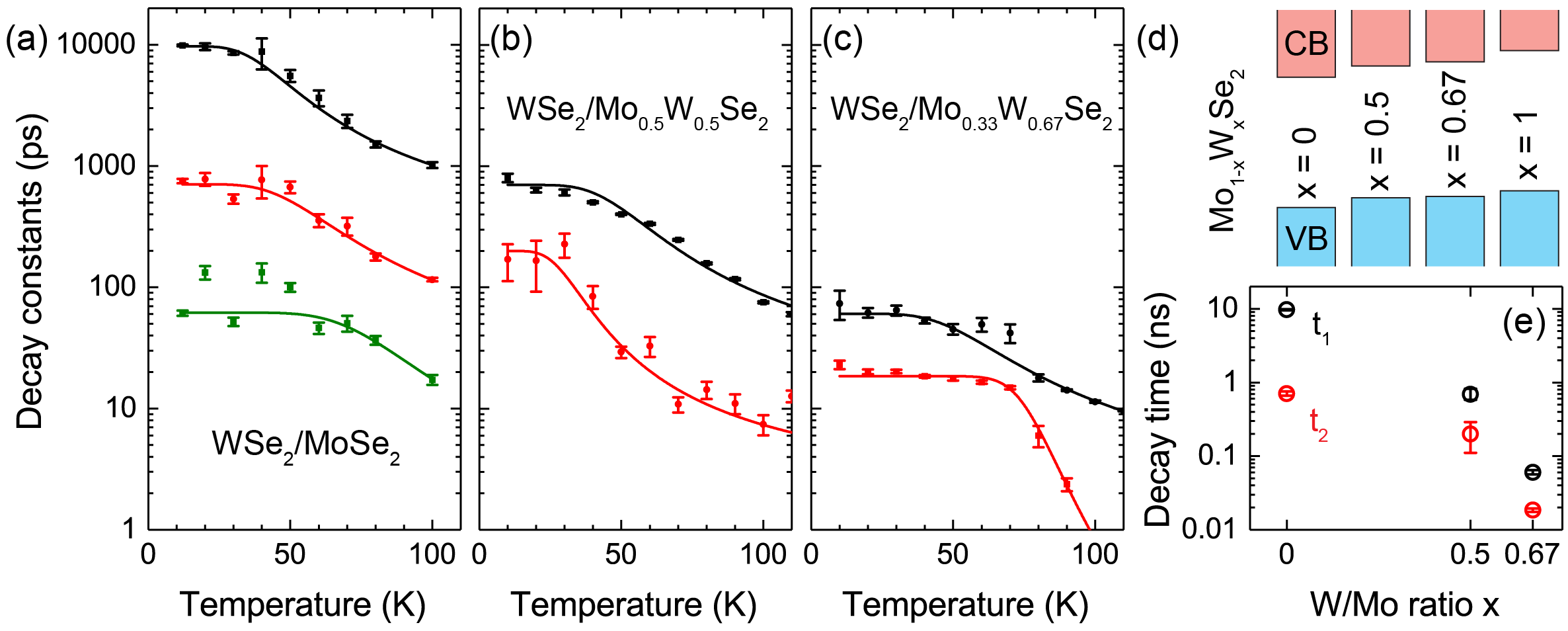}
	\caption{\textbf{Spin--valley dynamics in Mo$_{1-x}$W$_{x}$Se$_2$/WSe$_2$ heterobilayers with different alloy composition.} (a)-(c) Time constants, extracted from fits to measured Kerr rotation kinetics, as functions of sample temperature for heterostructures with different chemical composition (symbols), together with Arrhenius fits: MoSe$_2$/WSe$_2$ (a), Mo$_{0.5}$W$_{0.5}$Se$_2$/WSe$_2$ (b), and Mo$_{0.33}$W$_{0.67}$Se$_2$/WSe$_2$ (c). (d) Schematic representation of valence and conduction band edges in Mo$_{1-x}$W$_{x}$Se$_2$ monolayers with different W/Mo concentration ratio $x$. (e) Zero-temperature time constants extracted from Arrhenius fits for heterobilayers with different W/Mo concentration ratio in the alloy layer. Time constants associated with IX recombination are shown in black, and those for IX--e interaction are shown in red.}
	\label{fig:Composition}
\end{figure*}

We further explore the relation between the spin--valley dynamics in TMD heterobilayers and their band structure by studying TMD stacks based on alloys Mo$_{1-x}$W$_{x}$Se$_2$ with different W/Mo composition ratio $x$.
To this end, we measure KR kinetics in MoSe$_2$/WSe$_2$ ($x = 0$), Mo$_{0.5}$W$_{0.5}$Se$_2$/WSe$_2$ ($x = 0.5$), and Mo$_{0.33}$W$_{0.67}$Se$_2$/WSe$_2$ ($x = 0.67$) samples for varying sample temperature and extract spin--valley relaxation time constants from multi-exponential fits. 
For consistency, the alloy and WSe$_2$ monolayers in the selected heterobilayers are aligned, with corresponding stacking angles within few degrees.

The extracted time constants for the KR dynamics on the W resonance are plotted in Fig.~\ref{fig:Composition}a-c as functions of temperature for Mo$_{1-x}$W$_{x}$Se$_2$/WSe$_2$ heterobilayers with $x = 0$ (a), $x = 0.5$ (b), and $x = 0.67$ (c).
All of them exhibit thermally activated behavior similarly to the data shown in Fig.~\ref{fig:Temperature}c.
The corresponding Arrhenius fits are plotted as solid curves.
For the largest time constants (black), associated with IX recombination, we extract activation energies of 17~meV, 23~meV, and 22~meV for MoSe$_2$/WSe$_2$, Mo$_{0.5}$W$_{0.5}$Se$_2$/WSe$_2$, and Mo$_{0.33}$W$_{0.67}$Se$_2$/WSe$_2$ samples, respectively, demonstrating little variation with the W/Mo composition ratio $x$ and associated varying band edge energies~\cite{Zhang2019} as schematically illustrated in Fig.~\ref{fig:Composition}d.
This supports our assignment of thermally activated scattering on optical phonons as the underlying mechanism for the temperature dependencies observed for the W resonance in alloy-based HBLs.
Similarly, activation energies extracted for the smaller time constants (red), associated with IX--electron interaction, do not exhibit a clear trend in the W/Mo composition ratio $x$.

On the other hand, the extracted time constants show a clear and significant reduction with increasing W concentration in the alloy layer.
The extracted zero-temperature decay times are shown in Fig.~\ref{fig:Composition}e and decrease by an order of magnitude when the W/Mo composition ratio changes from $x = 0$ to $x = 0.5$ and by another order of magnitude when it changes to $x = 0.67$.
This behavior can be understood considering the evolution of the alloy ML band structure with parameter $x$ (Fig.~\ref{fig:Composition}d).
As the energy offsets between the VB and CB of the alloy and WSe$_2$ monolayers are reduced towards zero in the transition from a hetero- to a homo-bilayer, the increased hybridization between carrier wavefunctions in the aligned layers leads to faster IX recombination and more efficient IX--electron interaction.
We note that the observed lifetimes are limited by the population decay, whereas the degree of circular polarization can exhibit longer lifetimes even for homobilayers~\cite{Scuri2020}.
In future studies of alloy-based heterostructures, the valley dynamics can be extended by creating resident hole population through electrostatic control~\cite{Jin2018Science,Scuri2020}.

In summary, we have demonstrated that alloy-based TMD heterobilayers provide a powerful and tunable platform for studying valley physics in 2D semiconductor structures.
The valley dynamics in these systems is defined by the interactions between various excitonic species and resident carriers, and can be controlled via external parameters such as the stacking angle and chemical composition of the constituent layers.
Applying a simple kinetic model, we have revealed the role of valley-polarized resident electrons in the experimentally observed valley dynamics and separated the effects of interlayer exciton depolarization and recombination.
Both depolarization and recombination rates have been found to increase in transitions from a misaligned to aligned HBL, and from a hetero- to a homo-bilayer, effectively limiting the observed valley times.
Our results suggest alloying as a promising approach to control the valley polarization and its relaxation dynamics in 2D materials for the development of future valleytronic devices. \\




\noindent
{\bf \large Methods}\\
\noindent
\textbf{Sample fabrication.}
Samples of encapsulated Mo$_{1-x}$W$_{x}$Se$_2$/WSe$_2$ heterobilayers with different W/Mo concentration ratios $x = 0, 0.5, 0.67$ were prepared by dry transfer on Si substrates with a 90~nm thick thermal oxide.
Flakes of monolayer WSe$_2$, monolayer alloy Mo$_{1-x}$W$_{x}$Se$_2$, and few-layer hexagonal boron nitride (hBN) were mechanically exfoliated from commercial bulk crystals (HQ Graphene).
The stacking angle between the constituent monolayers was controlled during the transfer process and precisely measured using polarization-resolved second harmonic generation (see Supplementary Information).
The samples were characterized by reflectivity contrast (RC) and photoluminescence (PL) measurements performed at a temperature of 10~K.
The low-temperature RC and PL spectra for monolayers and heterobilayers are shown in the Supplementary Information.

\noindent
\textbf{Optical measurements.}
The spin--valley dynamics in hBN-encapsulated Mo$_{1-x}$W$_{x}$Se$_2$/WSe$_2$ heterobilayers was measured via time-resolved Kerr rotation micro-spectroscopy.
Frequency-degenerate pump and probe pulses are derived from a Ti:sapphire oscillator (Spectra Physics, Tsunami) and focused onto the sample with a long working distance microscope objective (Mitutoyo Plan Apo 20X/0.26).
The laser spot size on the sample is kept within a 5-10~\micron~range.
The polarization of the pump pulse is modulated at 50.1~kHz between the left-hand and right-hand circular polarization states with a photo-elastic modulator (Hinds Instruments, PEM-100), and the probe is polarized linearly perpendicular to the plane of incidence.
The pump and probe pulses are delayed with a mechanical delay line and offset spatially to yield a small (10$^\circ$) angular offset on the sample, which allows blocking the pump beam in the detection channel.
The rotation angle of the probe pulse polarization plane is measured in the reflection geometry using an optical bridge and lock-in detection at 50.1~kHz.
In order to reduce the noise due to the residual scattered pump beam, the probe beam is modulated at 450~Hz with an optical chopper, and the final Kerr rotation signal is detected with an additional lock-in amplifier.
For all time-resolved Kerr rotation measurements, the power of the pump beam is kept within the 50-200~$\mu$W range, and the probe-to-pump intensity ratio is kept at 1:4.
For all measurements, the samples are mounted in a low-vibration closed cycle cryostat (ColdEdge Technologies) and maintained at a controlled temperature in a 10-300~K range.\\




\noindent
{\bf \large Acknowledgements}\\
\noindent
The authors acknowledge funding from the Ministry of Education and Science of the Russian Federation through Megagrant No. 14.Y26.31.0015.
Time-resolved experiments were funded by the Russian Science Foundation, project No. 19-72-00146.
V.K. acknowledges support by the Government of the Russian Federation through the ITMO Fellowship and Professorship Program.
A.C. and A.I.T. thank the financial support of the European Union’s Horizon 2020 research and innovation programme under ITN Spin-NANO Marie Sklodowska-Curie grant agreement no. 676108.
A.G. and A.I.T. acknowledge funding by EPSRC (EP/P026850/1 and EP/S030751/1).
E.M.A. and A.I.T. thank the financial support of the Graphene Flagship Project under grant agreements 696656 and 785219.\\






\pagebreak
\widetext
\begin{center}
\textbf{\large Supplementary Information: Spin--valley dynamics in alloy-based transition metal dichalcogenide heterobilayers}
\end{center}
\setcounter{equation}{0}
\setcounter{figure}{0}
\setcounter{table}{0}
\setcounter{page}{1}
\makeatletter
\renewcommand{\thepage}{S\arabic{page}}
\renewcommand{\theequation}{S\arabic{equation}}
\renewcommand{\thefigure}{S\arabic{figure}}
\renewcommand{\bibnumfmt}[1]{[S#1]}
\renewcommand{\citenumfont}[1]{S#1}

\subsection*{Supplementary Note 1. Alloy-based heterobilayers.}
All heterobilayer samples studied in this work are assembled from high-quality large-area flakes exfoliated from bulk crystals.
The samples consist of WSe$_2$ and Mo$_{1-x}$W$_{x}$Se$_2$ monolayers encapsulated in 10-100~nm thick layers of hexagonal boron nitride (hBN).
The relative orientation of the layers inside a heterostructure is accurately determined from polarization-resolved second harmonic generation (SHG) measurements.
For SHG measurements, the samples are excited with $\sim 900$~nm pulses of $\sim 100$~fs duration derived from a Ti:sapphire oscillator, which generate SHG response at $\sim 450$~nm center wavelength.
The excitation pulses are linearly polarized, and the SHG signals are detected in the same polarization using a dichroic mirror and a spectrometer with CCD, where integration is performed over the entire SHG spectrum.

\begin{figure}[b]
	\includegraphics[width=\columnwidth]{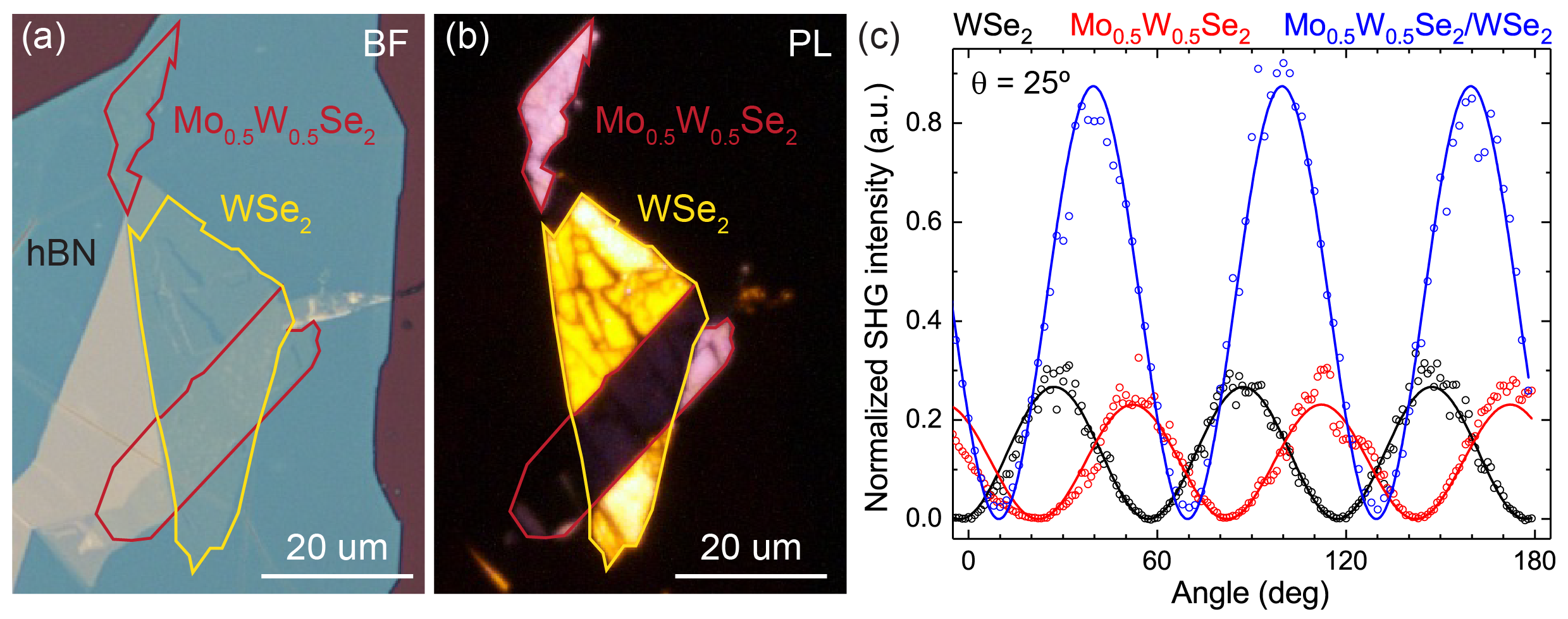}
	\caption{(a) Bright-field optical microscope image of a twisted Mo$_{0.5}$W$_{0.5}$Se$_2$/WSe$_2$ heterobilayer, showing monolayer WSe$_2$ (yellow) and Mo$_{0.5}$W$_{0.5}$Se$_2$ (red) regions. (b) Photoluminescence (PL) image of the sample, showing suppressed PL signal in the heterobilayer region due to fast charge separation between the two layers. (c) Polarization dependence of the second harmonic generation (SHG) signal for the WSe$_2$ monolayer (black), Mo$_{0.5}$W$_{0.5}$Se$_2$ monolayer (red), and heterobilayer (blue) regions on the sample, with an extracted twist angle of $\sim 25^{\circ}$.}
	\label{fig:SampleTwisted}
\end{figure}

The SHG intensity dependencies on the angle of sample rotation are measured for 3 different locations on the sample, including both WSe$_2$ and Mo$_{1-x}$W$_{x}$Se$_2$ monolayers and heterobilayer regions.
Each dependence is fitted with $I_\mathrm{SHG} = I_0 \operatorname{cos}^2(3\alpha - 3\alpha_0)$, where $\alpha_0$ represents the orientation of each layer.
Fig.~\ref{fig:SampleTwisted} shows an example of a twisted Mo$_{0.5}$W$_{0.5}$Se$_2$/WSe$_2$ heterobilayer, with a bright-field optical microscope image in (a), photoluminescence image in (b), and SHG polarization dependencies in (c), where SHG signals for the 1L WSe$_2$, 1L Mo$_{0.5}$W$_{0.5}$Se$_2$, and heterobilayer regions are shown with black, red, and blue colors, respectively.
The twist angle $\theta = \alpha^\mathrm{WSe2}_0 - \alpha^\mathrm{alloy}_0$ is estimated from fits as $\theta = 24.9 \pm 0.3^\circ$.

\begin{figure}[b]
	\includegraphics[width=\columnwidth]{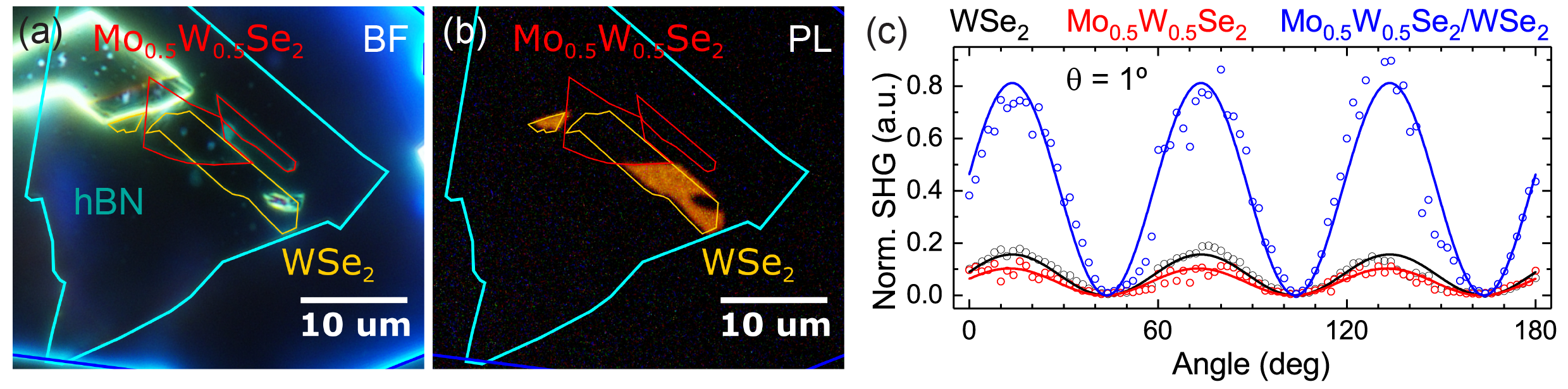}
	\caption{(a) Bright-field optical microscope image of an aligned Mo$_{0.5}$W$_{0.5}$Se$_2$/WSe$_2$ heterobilayer. (b) Photoluminescence image of the sample. (c) Polarization dependence of the SHG signal for the WSe$_2$ monolayer (black), Mo$_{0.5}$W$_{0.5}$Se$_2$ monolayer (red), and heterobilayer (blue) regions on the sample, with an extracted twist angle of $\sim 1^{\circ}$.}
	\label{fig:SampleAligned}
\end{figure}

An example of an aligned Mo$_{0.5}$W$_{0.5}$Se$_2$/WSe$_2$ heterobilayer is shown in Fig.~\ref{fig:SampleAligned}, with bright-field (a) and PL (b) images and SHG polarization dependencies (c).
A small twist angle of $\theta = 0.9 \pm 0.6^\circ$ is extracted from fits.
We note that the heterobilayer SHG intensity in this case is $\sim 4$~ times higher than for the monolayers, as expected for an aligned sample, allowing us to distinguish angles close to either $0^\circ$ or $60^\circ$.

\begin{figure}[t]
	\includegraphics[width=0.75\columnwidth]{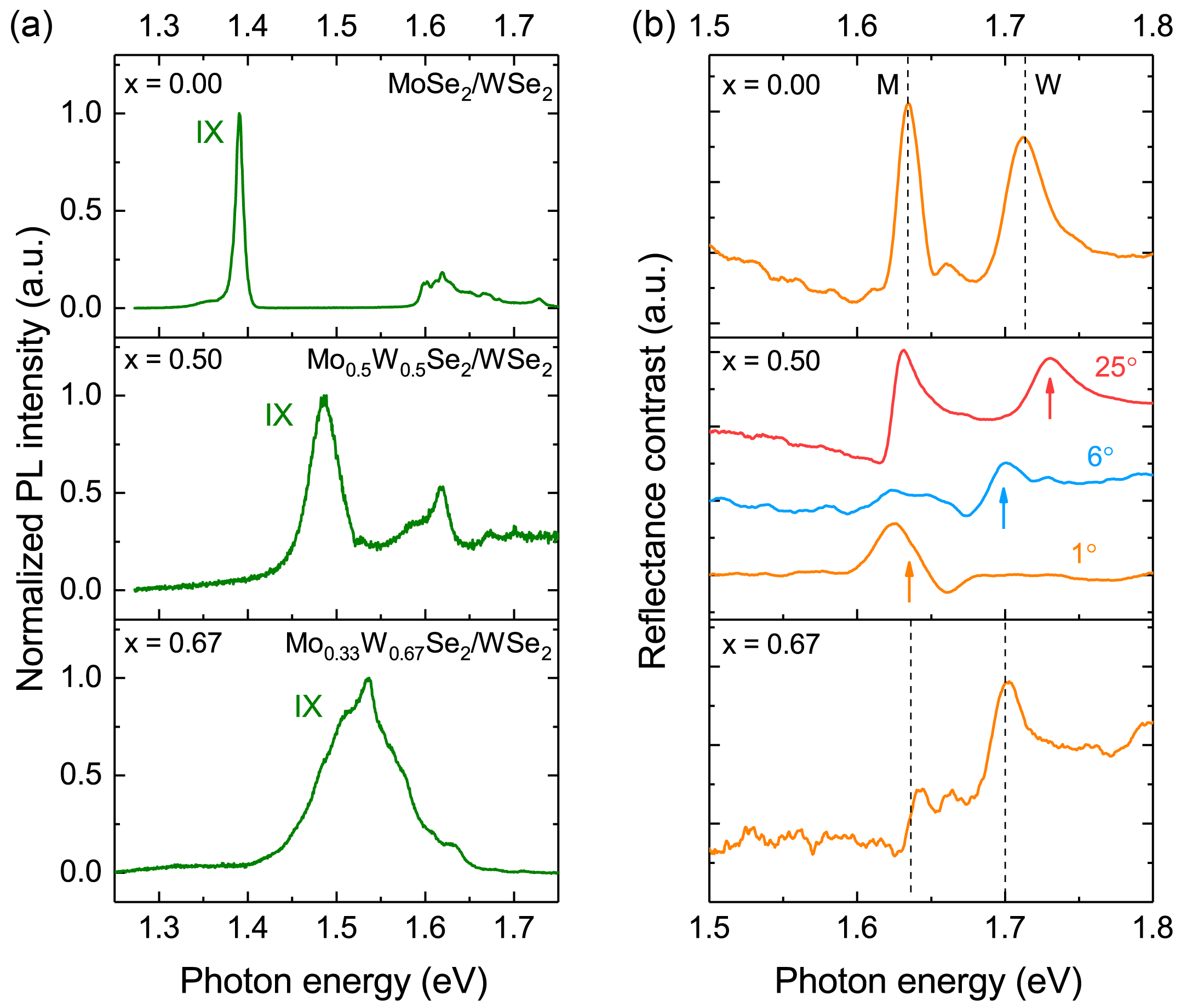}
	\caption{(a) Normalized PL spectra taken at 10~K on different heterobilayer samples: MoSe$_2$/WSe$_2$ (upper panel), Mo$_{0.5}$W$_{0.5}$Se$_2$/WSe$_2$ (middle panel), and Mo$_{0.33}$W$_{0.67}$Se$_2$/WSe$_2$ (lower panel). The interlayer exciton peak is labeled IX. The smaller peaks at higher frequencies are due to direct excitons, trions, and localized states in WSe$_2$ and alloy monolayers.}
	\label{fig:PLandRC}
\end{figure}

Fig.~\ref{fig:PLandRC} shows photoluminescence and reflectance contrast (RC) spectra measured at low temperature on different samples.
As observed in the PL spectra (a), the interlayer exciton PL peak (labeled IX) position shifts to higher energies when the relative W/Mo composition in the alloy monolayer increases from $x = 0$ (upper panel) to $x = 0.5$ (middle panel) and to $x = 0.67$ (lower panel), which is expected for a transition from a heterobilayer to a homobilayer.
Additionally, the interlayer exciton PL peak broadens, possibly due to the increased hybridization in the heterostructures with large values of $x$.
Fig.~\ref{fig:PLandRC}b shows the corresponding reflectance contrast spectra for samples with different W/Mo composition.
In the middle panel, RC spectra are shown for Mo$_{0.5}$W$_{0.5}$Se$_2$/WSe$_2$ samples with different twist angles.
As observed in the figure, the higher-energy resonance (labeled W, indicated with an arrow) redshifts for smaller twist angles, which is consistent with the Kerr rotation data presented in Fig.~4 of the main text.

\subsection*{Supplementary Note 2. Time resolved Kerr rotation micro-spectroscopy.}
\begin{figure}[t]
	\includegraphics[width=0.8\columnwidth]{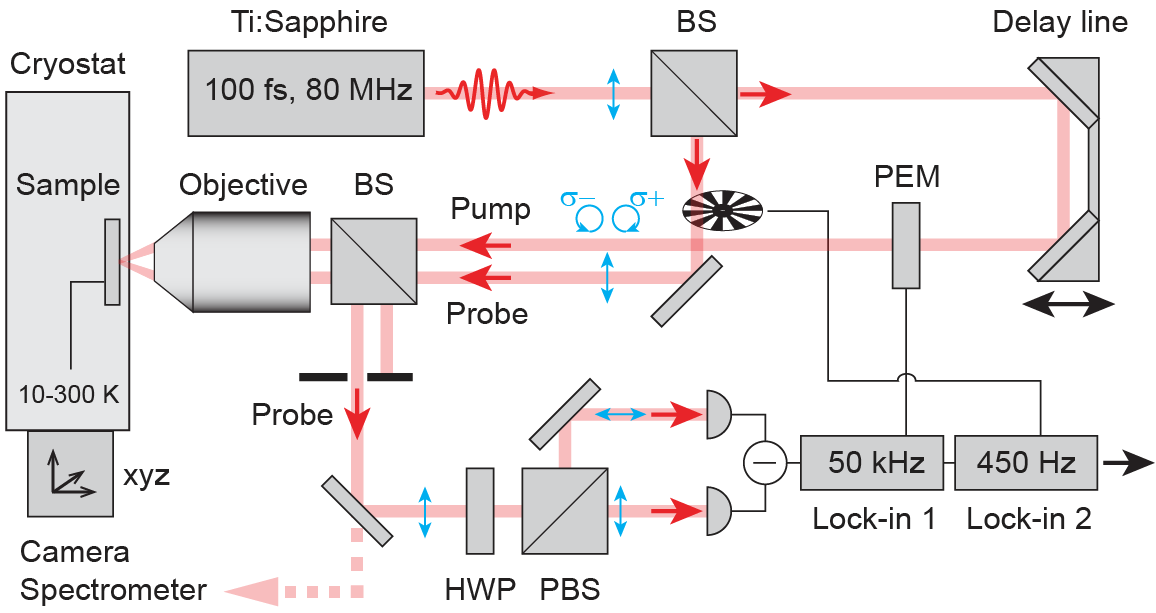}
	\caption{Schematic of the experimental setup for time-resolved Kerr rotation micro-spectroscopy of 2D heterobilayers. See description in the Methods section of the main text and Supplementary Note 2. BS: beamsplitter; PEM: photo-elastic modulator; HWP: half-wave plate; PBS: polarizing beamsplitter.}
	\label{fig:SupplementSetup}
\end{figure}

A schematic of the experimental setup for measuring Kerr rotation (KR) time traces is shown in Fig.~\ref{fig:SupplementSetup}.
In order to probe the small ($\mu$rad to mrad) KR angles observed in two-dimensional semiconductors, we modulate the pump beam between the left- and right-circularly polarized states with a photo-elastic modulator (PEM) at $\sim 50$~kHz frequency and use an optical bridge, balanced photodiode, and lock-in detection to obtain the KR signal. 
To achieve the micron-scale spatial resolution required to probe the dynamics in heterobilayers assembled from exfoliated WSe$_2$ and Mo$_{1-x}$W$_{x}$Se$_2$ monolayer flakes, we focus the pump and probe beams onto the sample with a 20X/0.26 long working distance microscope objective.
The degenerate pump and probe signals are separated spatially, with an offset between the parallel beams before the objective, which transforms in a slight angular offset at the sample.
The pump beam is blocked in the detection channel; however, due to scattering on the sample surface, some residual pump signal is still present at the detector.
In order to completely eliminate the contribution from the pump beam, we use a chopper in the probe beam at $\sim 450$~Hz and perform a secondary lock-in detection at that frequency to obtain the final KR values.

\subsection*{Supplementary Note 3. Theoretical modelling of spin--valley dynamics in heterobilayers.}
\subsubsection{Extraction of direct exciton parameters from the reflection spectra}
\begin{figure}[t]
	\includegraphics[width=1.1\textwidth]{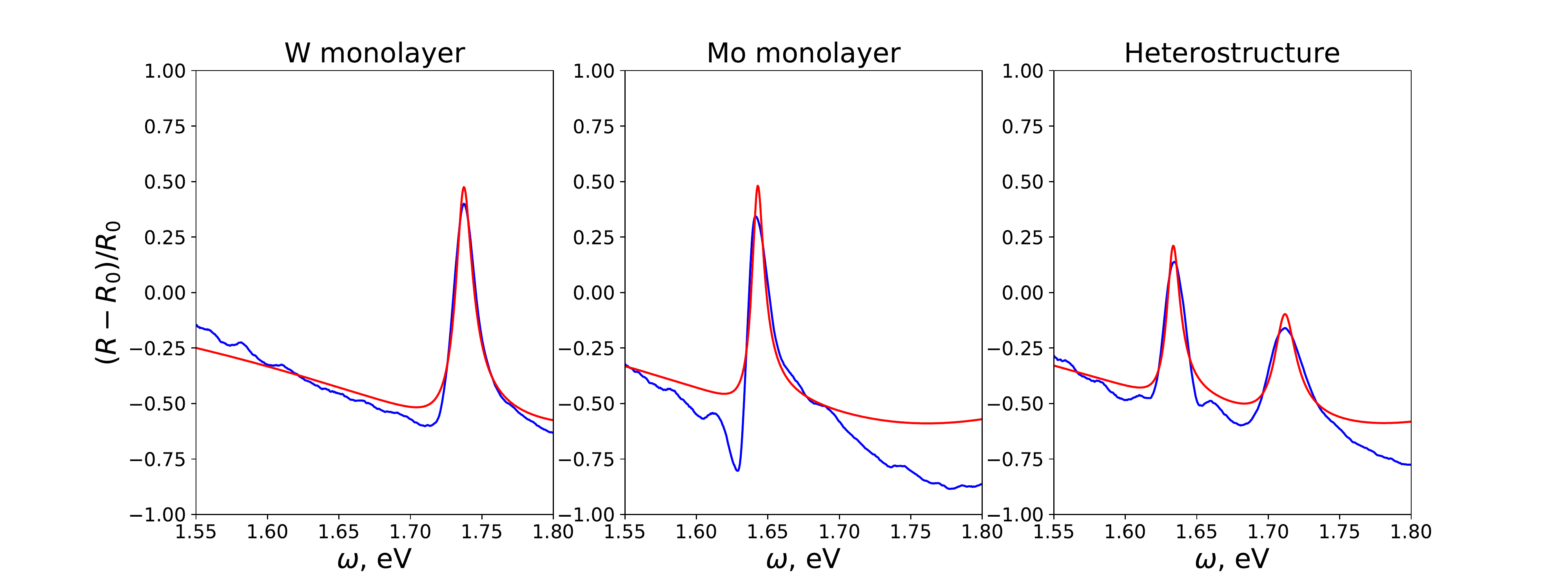}
	\caption{Experimental (blue) and fitted (red) reflectivity spectra for a MoSe$_2$/WSe$_2$ heterobilayer.}
	\label{fig:FanoFits}
\end{figure}

We first assume that the optical response of the probe is directly defined only by the direct excitons in the bilayer structure. Namely, the reflection coefficient of the bilayer is defined by the surface ac conductivity 
\begin{align}
    \sigma_{BL}(\omega)= \frac{-i\Gamma_{0W}}{\omega_{0W}-w-i(\Gamma_{0W}+\Gamma_W)}+\frac{-i\Gamma_{0M}}{\omega_{0M}-w-i(\Gamma_{0M}+\Gamma_M)}, r_{BL}=\frac{\sigma}{1+\sigma}
\end{align}
where $W,M$ correspond to direct excitons in the corresponding monolayers.
In order to account for the surrounding photonic structure, we derive the full reflection coefficient $r$ using the transfer matrix theory
\begin{align}
    r=\frac{T_{11}+T_{12}n_{Si}-T_{21}-T_{22}n_{Si}}{T_{11}+T_{12}n_{Si}+T_{21}+T_{22}n_{Si}},
\end{align}
and absolute value of reflection coefficient is $R=|r|^2$. Here  $T_{ij}$ are the components of the full transfer matrix of the system:
\begin{align}
    T=T_{hBN}\times T_{bilayer} \times T_{hBN}\times T_{SiO2},
\end{align}
Transfer matrix of the bilayer is given by
\begin{align}
    T_{bilayer}= \begin{pmatrix} 1 & 0 \\ \frac{-r_{BL}}{1+r_{BL}} & 1
    \end{pmatrix}
\end{align}
The generic spectrum of reflectivity can be fitted by two Fano lineshapes as shown below.
For the MoSe$_2$/WSe$_2$ sample the fitting results are shown in Fig.~\ref{fig:FanoFits}. 

The fitted parameters are
\begin{center}
\begin{tabular}{ |c|c| } 
 \hline
 $\omega_{X,W}$ & 1.711 eV  \\ 
 \hline
 $\omega_{X,Mo}$ & 1.633 eV \\ 
 \hline
 $\Gamma_{0,W}$ & 0.68 meV  \\
 \hline
 $\Gamma_{0,Mo}$ &  0.62 meV\\
 \hline 
 $\Gamma_{W}$ & 10.5 meV  \\
 \hline
 $\Gamma_{Mo}$ &  4.9 meV\\
 \hline 
 $d_{hBN,up}$ &  90 nm\\
 \hline
 $d_{hBN,down}$ &  35 nm \\
 \hline
\end{tabular}
\end{center}
The extracted parameters are then used to model the dynamics of the Kerr rotation signal.

\subsubsection{Interlayer tunneling rate estimation from ab-initio modelling}
The general expression for the interlayer tunneling current density between layers 1 and 2 reads~\cite{harrison1961tunneling}
\begin{align}
    j=\frac{4\pi e}{\hbar}\sum_{\mathbf{k}}\int dE |M_{1,2}(\mathbf{k},E)|^2\delta(\epsilon_1(\mathbf{k})-E)\rho_2(E)(f_1(E)-f_2(E)),
\end{align}
where $M_{1,2}$ is the tunneling matrix element between layers and $\rho_2(E)$ - is the density of states in the second layer. We can see that the tunneling current is roughly proportional to the product of the total concentration of the carriers in the first layer and the density of states in the second layer. The matrix element crucially depends on the mutual orientation of the layers~\cite{ruiz2019interlayer} and will play the role of the fitting parameter in our model. 

We have performed the ab-initio modelling of the density of states at the vicinity of the conduction and valence band edges for MoSe$_2$ and WSe$_2$. The modelling revealed, that the density of states at the conduction band edge is approximately 7 times larger than at the valence band edge. Thus, the interlayer hopping of electrons in the case of the $W$ resonance will be an order of magnitude more effective than in the case of the $Mo$ resonance. According to theoretical modelling~\cite{ruiz2019interlayer} and experimental measurements~\cite{jin2018ultrafast}, the characteristic tunneling times lie in the range $0.1-1$ ps. In the modelling we thus fix the ratio of the tunneling, and use one of the tunneling times as the fitting parameter.

\subsubsection{Model of the Kerr rotation}
To model the Kerr rotation angle, we assume that
that  the nonlinear effect results in the slight addition to reflection coefficient of circularly polarized waves. The Kerr rotation angle is then given by:
\begin{align}
    \Delta \theta = -\mathrm{Im}\left [\frac{\delta r_+ - \delta r_-}{2r}\right],
\end{align}
where $\delta r_{\pm}$ are the change in the reflection coefficients for the circularly polarized field of the corresponding helicity. We assume that the reflection signal in the vicinity of the exciton resonance is dominated by the direct exciton. Then, in the limit of the weak nonlinearity, the difference in reflection can be linearized as
\begin{align}
  \delta r_+ - \delta r_- = \frac{\partial r}{\partial \omega_0} (\omega_{0+}-\omega_{0-}) +  \frac{\partial r}{\partial \Gamma_0} (\Gamma_{0+}-\Gamma_{0-}) +  \frac{\partial r}{\partial \Gamma} (\Gamma_+-\Gamma_-)
\end{align}
Actually, this linearization should be performed separately for both direct excitons in each layer.
From the analysis of the fitted reflection and transmission coefficients, we  account only for the contribution of the shift of the resonant frequency, since other contributions appear to be negligibly small.

The frequency shift of the direct excitons happens due to the Coulomb interactions of the direct excitons with the direct and indirect excitons and with the residual carriers (electrons in our case). The Coulomb interaction crucially depends on the spin projections of the scattering particles. It is known that the dominant process is the scattering between the excitons with the same spin projections. The same holds for the exciton-electron interaction: the interaction is much larger for the case when the free and confined electrons have the same spin projections.
The exciton frequency shift can thus be written as
\begin{align}
    \omega_{0+}-\omega_{0-}=\delta\omega_0=V_{XX}^{DD}\tilde{n}_D m_{D}+V_{XX}^{DI}\tilde{n}_{I}m_{I}+V_{Xc}\tilde{n}_{c}m_{c},
\end{align}
where $\tilde{n}_{D,I,c}=(n^{+1}_{D,I,c}+n^{-1}_{D,I,c})/2$, $m_{D,I}=(n^{+1}_{D,I,c}-n^{-1}_{D,I,c})/\tilde{n}_{D,I,c}$ and $V$'s - are the corresponding matrix elements.

We now consider the cases of the W and Mo resonances separately.

\textit{W resonance}. We assume that there are no residual carriers at the W layer, so the exciton energy shift at the probe pulse are provided only by the direct exciton-exciton interaction $V_{XX}^{DD}$ and the interaction of excitons in W layer with the indirect excitons which are formed due to the tunnelling of the electrons to Mo layer shortly after the pump pulse, $V_{XX}^{DI}$.
There are two distinctive time scales, defining the spin polarization in the system. Within the first, short time-scale of the order of 1~ps, the processes which define the dynamics are: the direct exciton decay and depolarization, and electron tunneling to the Mo layer with the formation of interlayer excitons. The direct exciton energy shift for the times less than 10 ps can be evaluated as
\begin{align}
    \delta\omega_0=V_{XX}^{DD}n_{X0}e^{-(\gamma_t+\gamma_0+\gamma_{DP})t}+V_{XX}^{DI}n_{X0}\frac{\gamma_t}{\gamma_0+\gamma_t+\gamma_{DP}}(1-e^{-(\gamma_t+\gamma_0-\gamma_{DP})t}).
\end{align}
where $\gamma_t,\gamma_0$ and $\gamma_{DP}$ are the tunneling rate, exciton radiative lifetime and spin-valley depolarization time respectively. We note also, that due to the delocalization of the interlayer excitons, the matrix element of interaction $V_{XX}^{DI}>V_{XX}^{DD}$. This can lead to the increase of the Kerr signal at the time delays substantially larger than the pulse widths, provided that tunneling is the most efficient process in the system. In experiment we observe the increase of the Kerr rotation signals for the times $< 2$ ps for the case of $W$ resonance pump.

At longer delays, most  of the direct excitons have either decayed radiatively or formed the indirect excitons. At these times, the interaction of the indirect excitons with the residual electrons in the Mo layer takes place. Due to the exciton-electron interactions the net polarization is transferred from the indirect exciton to the electron subsystem. This dynamics is governed by the equations
\begin{align}
    &\dot{m}_c=-\tilde{n}_{IX,0}e^{-\gamma_{IX}t}V_{XC}^{IE}m_c(t)+m_{IX}(t)\tilde{n}_cV_{XC}^{IE}, \label{me}\\
    &\dot{m}_{IX}(t)=-\tilde{n}_cV_{XC}^{IE}m_{IX}(t)+\tilde{n}_{IX,0}e^{-\gamma_{IX}t}V_{XC}^{IE}m_c(t).\label{mix}
\end{align}
We also assume that initially the indirect excitons are fully polarized, $m_{IX}(0)=2$ and residual electron subsystem is completely depolarized, $m_e(0)=0$,
supplemented with initial conditions $m_e(0)=0,m_{IX}(0)=2$ . $V_{XC}^{IE}$ is the matrix element of the electron-indirect exciton interaction corresponding to the process for which the confined and free electron change spin projections after the scattering. The concentration of residual carriers $n_c$ is constant in time. We introduce $\gamma_{XC}=n_cV_{XC}^{IE}$ - the effective spin cross-correlation between the exciton and electron subsystems. We thus get a linear system of differential equations with time-dependent coefficients. The solution of this system for $m_{IX}\tilde{n}_{IX}(t)$ is given by
\begin{align}
   & m_{IX}=2-2\gamma_{XC}\exp[\tilde{n}_{IX}(t)V_{XC}/\gamma_{IX}]\int_0 ^t dt' e^{\gamma_{XC}(t'-t)}\exp[-\tilde{n}_{IX}(t')V_{XC}/\gamma_{IX}]
\end{align}
for $\gamma_{IX}\ll \gamma_{XC}$ the time dependence can be separated at the short time-scale where the magnetization decays as 
$\exp[-2\gamma_{XC}t]$ and long time scale when the magnetization decays as $\exp[-\gamma_{IX}t]$. The observed intermediate relaxation times of the order of 100-1000 ps, correspond to $\gamma_{XC}\approx 1 \mu$eV.
Moreover the shape of the Kerr-angle time dependence allows to evaluate the residual electron concentration and the exchange matrix element separately. Namely, for the Mo$_{0.5}$W$_{0.5}$Se$_2$/WSe$_2$ sample the concentration of residual carriers is approximately $10^{11}~\mathrm{cm}^{-2}$ and the matrix element is $V_{XC}^{IE}\approx 10^{-2}~\mu\mathrm{eV}~{\mu}m^2$.

The exciton-electron exchange interaction supported by the spin flip is accompanied by the intervalley scattering and thus requires the emission or absorption of the phonon providing the momentum conservation. The temperature dependence of the depolarization rate thus should exhibit the step like behaviour reflecting the temperature induced activation of the phonon absorption process at the thermal energies close to the phonon energy. In experiments we observe the step like behaviour of the depolarization time with the activation energy on the order of $30$ meV which corresponds to the characteristic phonon energies in MoSe$_2$~\cite{mobaraki2019temperature}.

\textit{Mo resonance}. The situation is qualitatively different, when we pump the Mo layer.
First, the tunneling of holes is far less efficient and happens with characteristic time about 2 ps. Then, in MoSe$_2$ depolarization of the direct excitons happens at shorter time scales of the order of 500 fs. At the same time, when we pump resonantly the Mo layer, the spin polarization of the residual carriers is induced due to the trion formation with the subsequent depolarization and decay~\cite{harmon2010theory}. Within this process, an electron with a defined spin projection is subtracted from the residual electrons for the trion formation, but upon the trion decay the released electron is depolarized, thus a net spin polarization of residual electrons remains. 
At longer times the electron and indirect exciton subsystems exchange polarization analogously to Eq.~\eqref{me},~\eqref{mix}. Then upon the arrival of the probe pulse, the Kerr rotation is due to the shift of the exciton frequency due to the interaction of direct Mo exciton with indirect excitons and spin-polarized residual electrons.

\end{document}